\documentclass[twocolappendix, twocolumn]{aastex701}

\newcommand{\LP}[1]{\ensuremath{\mathrm{LP}_{#1}}}
\accepted{September 14, 2025}
\submitjournal{ApJL}

\begin{document}

\title{On-sky demonstration of subdiffraction-limited astronomical measurement using a photonic lantern}

\author[orcid=0000-0003-1392-0845,gname='Yoo Jung',sname='Kim']{Yoo Jung Kim}
\affiliation{Department of Physics \& Astronomy, University of California Los Angeles, Los Angeles, 475 Portola Plaza, Los Angeles, CA 90095, USA}
\email[show]{yjkim@astro.ucla.edu}  

\author[orcid=0000-0002-0176-8973,gname='Michael',sname='Fitzgerald']{Michael P. Fitzgerald}
\affiliation{Department of Physics \& Astronomy, University of California Los Angeles, Los Angeles, 475 Portola Plaza, Los Angeles, CA 90095, USA}
\email{mpfitz@ucla.edu}

\author[orcid=0000-0003-4018-2569,gname='Sebastien',sname='Vievard']{S\'ebastien Vievard}
\affiliation{Space Science and Engineering Initiative, College of Engineering, University of Hawai'i, 640 North Aohoku Place, Hilo, HI 96720, USA}
\affiliation{Subaru Telescope, National Astronomical Observatory of Japan, 650 North Aohoku Place, Hilo, 96720, HI, USA}
\affiliation{Institute for Astronomy, University of Hawai’i, 640 N. Aohoku Pl., Hilo, HI, 96720, USA}
\email{vievard@hawaii.edu}

\author[orcid=0000-0001-8542-3317, gname='Jonathan',sname='Lin']{Jonathan Lin}
\affiliation{Department of Physics \& Astronomy, University of California Los Angeles, Los Angeles, 475 Portola Plaza, Los Angeles, CA 90095, USA}
\email{jon880@astro.ucla.edu}

\author[orcid=0000-0002-6171-9081,gname='Yinzi',sname='Xin']{Yinzi Xin}
\affiliation{Department of Astronomy, California Institute of Technology, 1200 East California Boulevard, Pasadena, CA 91125, USA}
\email{yxin@caltech.edu}

\author[orcid=0000-0001-6341-310X,gname='Miles',sname='Lucas']{Miles Lucas}
\affiliation{Institute for Astronomy, University of Hawai’i, 640 N. Aohoku Pl., Hilo, HI, 96720, USA}
\affiliation{Subaru Telescope, National Astronomical Observatory of Japan, 650 North Aohoku Place, Hilo, 96720, HI, USA}
\affiliation{Steward Observatory, University of Arizona, 933 N Cherry Ave,Tucson, AZ 85719, USA}
\email{mlucas@hawaii.edu}

\author[orcid=0000-0002-1097-9908,gname='Olivier',sname='Guyon']{Olivier Guyon}
\affiliation{Subaru Telescope, National Astronomical Observatory of Japan, 650 North Aohoku Place, Hilo, 96720, HI, USA}
\affiliation{Steward Observatory, University of Arizona, 933 N Cherry Ave,Tucson, AZ 85719, USA}
\affiliation{College of Optical Sciences, University of Arizona, 1630 E University Blvd, Tucson, AZ 85719, USA}
\affiliation{Astrobiology Center, 2 Chome-21-1, Mitaka, 181-8588, Japan}
\email{guyon@naoj.org}

\author[orcid=0000-0002-3047-1845,gname='Julien',sname='Lozi']{Julien Lozi}
\affiliation{Subaru Telescope, National Astronomical Observatory of Japan, 650 North Aohoku Place, Hilo, 96720, HI, USA}
\email{lozi@naoj.org}

\author[orcid=0000-0003-4514-7906,gname='Vincent',sname='Deo']{Vincent Deo}
\affiliation{Optical Sharpeners SAS, Manosque, France}
\affiliation{Subaru Telescope, National Astronomical Observatory of Japan, 650 North Aohoku Place, Hilo, 96720, HI, USA}
\email{vdeo@naoj.org}

\author[orcid=0000-0002-1342-2822,gname='Elsa',sname='Huby]{Elsa Huby}
\affiliation{LIRA, Observatoire de Paris, Universit´e PSL, Sorbonne Universit´e, Universit´e Paris Cit´e, CY Cergy Paris Universit´e, CNRS, 5 place Jules Janssen, Meudon, 92195, France}
\email{elsa.huby@obspm.fr}

\author[orcid=0000-0002-6948-0263,gname='Sylvestre',sname='Lacour']{Sylvestre Lacour}
\affiliation{LIRA, Observatoire de Paris, Universit´e PSL, Sorbonne Universit´e, Universit´e Paris Cit´e, CY Cergy Paris Universit´e, CNRS, 5 place Jules Janssen, Meudon, 92195, France}
\email{sylvestre.lacour@obspm.fr}

\author[orcid=0009-0002-0929-0588,gname='Manon',sname='Lallement']{Manon Lallement}
\affiliation{LIRA, Observatoire de Paris, Universit´e PSL, Sorbonne Universit´e, Universit´e Paris Cit´e, CY Cergy Paris Universit´e, CNRS, 5 place Jules Janssen, Meudon, 92195, France}
\email{Manon.LALLEMENT@obspm.fr}

\author[orcid=0009-0007-8546-4363,gname='Rodrigo',sname='Amezcua-Correa]{Rodrigo Amezcua-Correa}
\affiliation{The College of Optics and Photonics, University of Central Florida, 4304 Scorpius Street, Orlando, FL 32816, USA}
\email{r.amezcua@creol.ucf.edu}

\author[orcid=0000-0002-5606-3874,gname='Sergio',sname='Leon-Saval']{Sergio Leon-Saval}
\affiliation{Sydney Institute for Astronomy, The University of Sydney, School of Physics A28, Sydney, NSW 2006, Australia}
\email{sergio.leon-saval@sydney.edu.au}

\author[orcid=0000-0002-8352-7515,gname='Barnaby',sname='Norris']{Barnaby Norris}
\affiliation{Sydney Institute for Astronomy, The University of Sydney, School of Physics A28, Sydney, NSW 2006, Australia}
\email{barnaby.norris@sydney.edu.au}

\author[gname='Mathias',sname='Nowak']{Mathias Nowak}
\affiliation{LIRA, Observatoire de Paris, Universit´e PSL, Sorbonne Universit´e, Universit´e Paris Cit´e, CY Cergy Paris Universit´e, CNRS, 5 place Jules Janssen, Meudon, 92195, France}
\email{mathias.nowak@obspm.fr}

\author[orcid=0000-0001-6871-6775,gname='Steph',sname='Sallum']{Steph Sallum}
\affiliation{Department of Astronomy \& Astrophysics, University of California Santa Cruz, 1156 High Street, Santa Cruz, CA 95064, USA}
\email{ssallum@ucsc.edu}

\author[orcid=,gname='Jehanne',sname='Sarrazin']{Jehanne Sarrazin}
\affiliation{LIRA, Observatoire de Paris, Universit´e PSL, Sorbonne Universit´e, Universit´e Paris Cit´e, CY Cergy Paris Universit´e, CNRS, 5 place Jules Janssen, Meudon, 92195, France}
\email{jehanne.sarrazin@obspm.fr}

\author[gname='Adam',sname='Taras']{Adam Taras}
\affiliation{Sydney Institute for Astronomy, The University of Sydney, School of Physics A28, Sydney, NSW 2006, Australia}
\email{adam.taras@sydney.edu.au}

\author[gname='Stephanos',sname='Yerolatsitis]{Stephanos Yerolatsitis}
\affiliation{The College of Optics and Photonics, University of Central Florida, 4304 Scorpius Street, Orlando, FL 32816, USA}
\email{stephanos.yerolatsitis@ucf.edu}

\author[orcid=0000-0001-5213-6207,gname='Nemanja',sname='Jovanovic']{Nemanja Jovanovic}
\affiliation{Department of Astronomy, California Institute of Technology, 1200 East California Boulevard, Pasadena, CA 91125, USA}
\email{nem@caltech.edu}

\begin{abstract} 

Resolving fine details of astronomical objects provides critical insights into their underlying physical processes. This drives in part the desire to construct ever-larger telescopes and interferometer arrays and to observe at shorter wavelength to lower the diffraction limit of angular resolution. Alternatively, one can aim to overcome the diffraction limit by extracting more information from a single telescope's aperture. A promising way to do this is spatial mode-based imaging, which projects focal-plane field onto a set of spatial modes before detection, retaining focal-plane phase information crucial at small angular scales but typically lost in intensity imaging. However, the practical implementation of mode-based imaging in astronomy from the ground has been challenged by atmospheric turbulence. Here, we present the first on-sky demonstration of a subdiffraction-limited, mode-based measurement using a photonic lantern (PL)-fed spectrometer installed on the SCExAO instrument at the Subaru Telescope. We introduce a novel calibration strategy that mitigates time-varying wavefront error and misalignment effects, leveraging simultaneously recorded focal-plane images and using a spectral-differential technique that self-calibrates the data. Observing the classical Be star $\beta$ CMi, we detected spectral-differential spatial signals and reconstructed images of its H$\alpha$-emitting disk. We achieved an unprecedented H$\alpha$ photocenter precision of $\sim50$\,$\mu$as in about 10-minute observation with a single telescope, measuring the disk's near-far side asymmetry for the first time. This work demonstrates the high precision, efficiency, and practicality of photonic mode-based imaging techniques to recover subdiffraction-limited information, opening new avenues for high angular resolution spectroscopic studies in astronomy.

\end{abstract}



\section{Introduction}\label{sec:intro} 

Many areas of astrophysical study require measuring the distribution of light at very high angular resolution. The sharpest angular resolutions in astronomy are achieved by long-baseline interferometry, probing the smallest scales of the universe, such as the innermost vicinity of supermassive black holes \citep{gra18, eht19}, innermost regions of circumstellar disks \citep{gra20, gra24}, and stellar surfaces \citep{monnier07}. However, the technical challenge of long-baseline interferometry increases dramatically at shorter optical wavelengths, where atmospheric phase errors are more severe and mechanical stability requirements are more demanding.

A complementary approach to achieving higher angular resolution is through the development of techniques that extract more information from a single telescope aperture. This path is based on the principle that the Rayleigh criterion for diffraction-limited resolution is not fundamental, but rather a consequence of sampling intensity of the focal-plane field imaged on a detector that discards phase information about the incoming light \citep{tsang16}. One of the ways of accessing this lost phase information is spatial mode-based measurement, which has theoretical foundations in quantum information theory \citep{tsang16, rehacek17a, tsang19}. When the angular scale of an object approaches the diffraction limit of the imaging system, the focal-plane field can be described by a small number of spatial modes. Therefore, rather than sampling the focal-plane intensity in detector pixel basis, one can instead measure the power after projecting the field onto a set of a finite number of spatial modes. The potential of this technique to achieve super-resolution has since been demonstrated in laboratory experiments \citep[e.g.,][]{tham17, boucher20, rouviere24}.

\begin{figure*}[hbt!]
\centering
\includegraphics[width=0.7\textwidth]{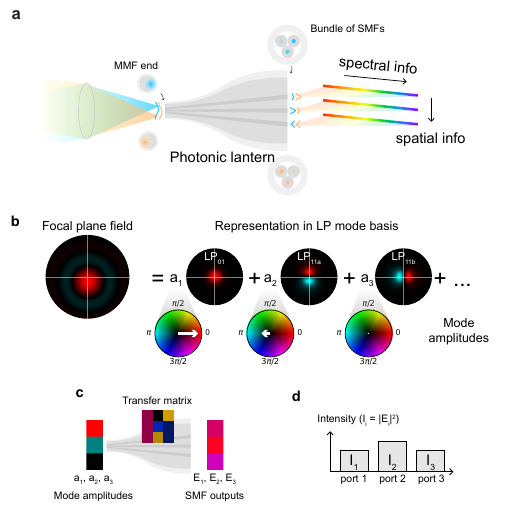}
\caption{
The PL-IFU concept. 
{\bf a.} A conceptual diagram of using a PL as a compact integral-field unit. The light collected by the telescope is coupled to the multimode end of the PL at the focal plane, which is then converted into multiple single-mode beams encoding the spatial information of the incoming wavefront. Using the spectrally dispersed single-mode outputs, both spectral and spatial information can be recovered at high angular resolution within a small field of view (few $\lambda/D$), enabling detailed characterization of compact astronomical sources. 
{\bf b.} An illustrative diagram of spatial mode decomposition of focal-plane field. Using linearly polarized (LP) mode basis that describes spatial modes of circular step-index optical fibers, a PSF with a small shift is primarily described as a linear combination of the fundamental \LP{01} and first-order \LP{11} modes, with complex-valued mode amplitudes. The complex mode amplitudes are represented as phasors (white arrows), where the arrow's length and angle correspond to the magnitude and phase of the coefficient. 
{\bf c.} A PL has unique complex-valued mixing coefficients (transfer matrix) that map the mode amplitudes to complex amplitudes of SMF outputs; for simplicity, we describe the case of a 3-port PL here.
{\bf d.} By imaging SMF outputs on a detector, one measures intensities of the SMF outputs, which have spatial information of the input.
}\label{fig:concept}
\end{figure*}

One possible implementation of spatial mode-based measurement employs a photonic lantern \citep[PL;][]{leon-saval05, birks15}, which is a type of gradually varying waveguide that has a multimode input and multiple single-mode outputs, efficiently converting multimode light into single-mode beams (Fig. \ref{fig:concept}a). When the PL's multimode input is placed at the telescope's focal plane, the focal-plane field is decomposed into the multimode input's mode basis (Fig. \ref{fig:concept}b), retaining low-order amplitude and phase information. These modal components are then coherently mixed within the PL to produce a set of single-mode outputs (Fig. \ref{fig:concept}c). The complex amplitude of each output is a linear combination of the input components, with the specific mixing coefficients being fixed by the design and construction of the PL. As a result, the relative intensities of the outputs --- the directly measurable quantities --- encode the spatial information of the input scene (Fig. \ref{fig:concept}d) \citep{kim24_imaging, kim24_sa, kim24_3PL, eikenberry24}. The single-mode output intensities can be measured spectrally by feeding a diffraction-limited spectrometer \citep{betters14, lin21}. This allows a PL-fed spectrometer to function as a compact and efficient integral field unit (IFU) for high angular resolution spectroscopy, within a small field of view.

However, ground-based observing conditions, especially atmospheric turbulence, hinder the practical implementation of spatial-mode-based measurements. Even with high-order adaptive optics (AO) \citep{guyon18} actively correcting wavefront errors (WFEs), residual tip-tilt jitter (rapid image motion) and unsensed aberration modes such as petal modes introduced by the low-wind effect \citep{ndiaye_lwe} further degrade the PSF quality. These low-order aberrations couple into PLs along with the true spatial signal, significantly hampering the measurement. It is necessary to develop dedicated observation and calibration strategies to isolate true spatial signals from aberrations.

In this Letter, we present the first on-sky demonstration of a high angular resolution, mode-based measurement using a photonic lantern-fed spectrometer. A critical aspect of the setup was the use of a hybrid imaging approach, in which infrared focal-plane images were synchronized with the visible PL output spectrum to monitor the PSF at the PL's multimode input. This allowed us to leverage the natural image motion from residual WFEs to spatially map PL responses by frame-sorting, essential for image reconstruction. Moreover, we used a spectral-differential method (e.g., spectroastrometry \citep{bailey98, whelan08}): using measurements at reference wavelengths to calibrate those at targeted wavelengths. This eliminates the need for separate calibrator observations, which can be compromised by misalignment and quasi-static aberrations. 

With this technique, we detected spectroastrometric signals from the H$\alpha$-emitting disk around the classical Be star $\beta$ CMi, measuring its near-far side asymmetry for the first time. In about 10 minutes of on-sky observation with a single telescope, we achieved photocenter precision of $\sim$50\,$\mu$as at visible wavelengths. This work opens a new window for high angular resolution spectroscopic studies with a single telescope.

\begin{figure*}[hbt!]
\centering
\includegraphics[width=1.0\textwidth]{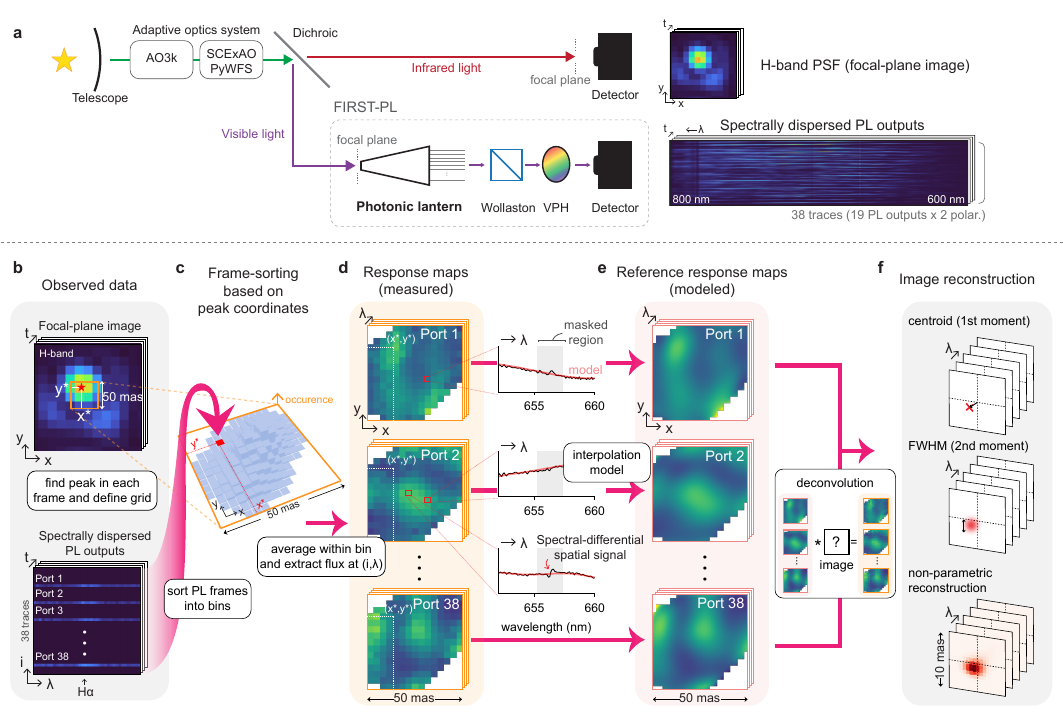}
\caption{
{The observation setup and data processing procedure.} 
{\bf a.} A simplified diagram of the observation setup. After two-stage AO correction of the light collected by the Subaru Telescope, the visible light is directed to the FIRST-PL module recording spectrally dispersed outputs of the 19-port PL, and the infrared light is sent to a high-speed detector at the focal plane with an $H$-band filter in. 
{\bf b.} Example $H$-band PSF and corresponding PL spectrum. The PSF peak location ($x^*, y^*$) is measured from each $H$-band focal-plane frame and is associated with the corresponding PL spectrum frame. 
{\bf c.} Frame sorting procedure. The PL spectrum data are binned by the PSF peak locations, with a resolution of $15\times 15$ spanning 48.6\,mas. 
{\bf d.} Reconstructed PL response maps. Each port's response map exhibits a unique response, with a gradual variation across wavelengths. The model responses are constructed by masking the H$\alpha$ wavelength range (where spatial signals are expected) and fitting the measured response maps with polynomial functions, then interpolating the coefficients in the masked wavelength range. 
Spectral-differential spatial signals are identified as deviations of the response maps from the model.
{\bf e.} Model reference response maps, which correspond to response maps of a point-like source.
{\bf f.} Image reconstruction is achieved by solving for the intensity distribution that, when convolved with the reference response maps, best describes the measured response maps. The first and second moments can be estimated by modeling the image as a point source and 2D Gaussian, respectively. Without any simplifying models non-parametric image reconstruction can be achieved. 
}\label{fig:method}
\end{figure*}

\section{Observations}\label{sec:obs} 

$\beta$ CMi is a well-studied late-type classical Be star (B8Ve). Classical Be stars have a decretion disk consisting of the materials ejected from the fast-rotating star \citep{rivinius_review}. Intense stellar radiation ionizes the disk and produces strong emission lines from recombining hydrogen, whose Doppler-shifted light from the approaching and receding sides appears as a double-peaked profile. Due to the compact characteristic size of the light emitting region (a few stellar radii), spatially resolving classical Be star disks is feasible only for the nearest targets, using optical long-baseline interferometry \citep{thom86, vakili98, delaa11, meilland12} or spectroastrometry \citep{oudmaijer08, wheelwright12}; for $\beta$ CMi, the examples include \cite{tycner05, kraus12, wheelwright12}.

We observed $\beta$ CMi ($V = 2.9$\,mag) on two engineering nights (2024 September 17 and 2025 February 11 UT) under proposal IDs S24B-EN03 and S25A-EN03 (PI: Julien Lozi), using the FIRST-PL module consisting of a 19-port photonic lantern feeding a mid-resolution spectrometer ($R\sim3000$, $\lambda = 0.6$--0.8\,$\mu m$) \citep{vievard24, firstpl_poster} installed on the Subaru Coronagraphic Extreme Adaptive Optics (SCExAO) instrument \citep{jovanovic15_scexao} at the 8.2\,m Subaru Telescope (Table \ref{tab:obslog}). 
A primary challenge for mode-based measurement is the need for a precise calibration of static and time-varying WFEs and misalignments. We addressed this by recording near-infrared point-spread function (PSF) simultaneously with the acquisition of the PL output spectrum. This ancillary PSF data was then used in the post-processing step as detailed in \S\ref{sec:pp}.

\subsection{The FIRST-PL module at Subaru/SCExAO}

FIRST-PL is installed as one of the visible modules of Subaru/SCExAO. Fig. \ref{fig:method}a shows the instrument setup. The focal-plane light is injected into the multimode end of the PL, while its outputs feed the mid-resolution spectrometer. Within the FIRST-PL spectrometer, the 19 single-mode outputs form a pseudo-slit, passed through a Wollaston prism to split orthogonal polarizations, and then spectrally dispersed by a volume phase holographic (VPH) grating. The resulting 38 spectral traces are imaged to the Hamamatsu ORCA-Quest qCMOS detector (hereafter FIRST-PL camera). The Wollaston prism, combined with a waveplate, enables differential polarimetry of each spectral trace. However, given the low degree of polarization expected from $\beta$ CMi ($\sim0.05\%$ \citep{klement15}), we ignored polarization and treated all the 38 output spectra separately. A focal ratio of f/8 was used, giving the optimal injection efficiency determined with lab experiments. See \cite{vievard24} for the details on the integration of the 19-port PL to the SCExAO instrument and in-lab characterization.

\subsection{Simultaneous observations with FIRST-PL and H-band PSF camera}

During both nights, the first-stage wavefront correction was performed using the recently upgraded AO3k \citep{lozi22_ao3k, lozi24_ao3k} and the second-stage wavefront correction was performed by the visible pyramid wavefront sensor that drove the SCExAO deformable mirror (DM). After correction by the SCExAO DM, the light is split with a dichroic beam splitter reflecting visible light ($\lambda<950$\,nm) to a separate optical bench, where the FIRST-PL module is located. The transmitted near-infrared light ($\lambda>950$\,nm) is distributed across different modules, and a 90:10 gray beam splitter was used to pick off 10\% of the near-infrared light for the SCExAO internal camera (FLI C-RED 2; hereafter PSF camera). The PSF camera was placed at the focal-plane with an $H$-band filter in, recording PSF images simultaneously with PL data acquisition (Fig. \ref{fig:method}a).

The FIRST-PL camera detector integration time (DIT) was set to 20~ms and 10~ms for the two observations, chosen to be short enough compared to the atmospheric decorrelation timescale after the two-stage AO correction while still providing sufficient signal per frame. The PSF camera integration time was chosen to run a few times faster than the FIRST-PL camera, 5.3~ms and 2.7~ms for the two observations. Timestamps associated with each frame were also recorded. This temporal oversampling of the PSF allows for more precise synchronization between the two data streams in post-processing using the timestamps.

The wavefront correction quality was variable during both nights, with significant PSF splitting due to the low-wind effect \citep{ndiaye_lwe}, which improved with time during the observation. Thus, we used the last 14 minutes out of 35 minutes of observation for the September 2024 data and the last 8 minutes out of 19 minutes for the February 2025 data that showed relatively stable wavefront. This resulted in about 50,000 frames for both observations, with similar $H$-band Strehl ratio ($\sim50$\%). 
Although nearly diffraction-limited, tip-tilt jitter (about 15\,mas root-mean-square (RMS), nearly one diffraction element at H$\alpha$ wavelength) and the low-wind effect \citep{ndiaye_lwe} dominated residual WFEs for both observations (Fig. \ref{fig:psf_extended}). These time-varying WFEs coupled into the PL, causing its output spectrum to fluctuate significantly.

\section{Post-processing}\label{sec:pp} 

\subsection{The response map method}

As discussed in \S\ref{sec:intro}, PL relative intensities encode spatial information of the input scene. For our target, $\beta$ CMi, we expect that the normalized spectra --- the spectrum of each port divided by the summed spectrum across all the ports --- to exhibit distinctive features across H$\alpha$ line, as the red- and blue-shifted light emitting regions are spatially displaced. However, when we averaged the normalized spectra over all observation frames, no such feature was detected (\S\ref{sec:a:PCA}) because time-varying WFEs from atmospheric turbulence dilute subtle spatial signals \citep{kim24_sa, kim24_3PL}.

To overcome this, we developed a new approach that utilizes the simultaneously recorded PSF frames. The PL frames are sorted into two-dimensional bins according to their corresponding ($x$, $y$) PSF peak position and then averaged within each bin (Fig. \ref{fig:method}c). This process builds a two-dimensional ``response map" for each port and wavelength channel (Fig. \ref{fig:method}d). These measured response maps, $M(\lambda,i)$, can be modeled as a convolution of the true astronomical image, ${\rm Image}(\lambda)$, with the response of a point source, which we term the ``reference response map", $M_0(\lambda,i)$ (Fig. \ref{fig:method}e): 
\begin{equation}\label{eq:model}
    M(\lambda, i) = {\rm Image}(\lambda) * M_0(\lambda, i).
\end{equation}
where $i$ is the port index and $*$ denotes the convolution.
Recovering the astronomical image thus becomes a deconvolution problem (Fig. \ref{fig:method}f). Photocenter shifts in the source appear as translations in $M$ relative to $M_0$, while spatially extended emission appears as broadening. 

This response map approach is powerful as it simultaneously addresses two key challenges in ground-based mode-based imaging \citep{kim24_3PL}: 1) it filters out tip-tilt jitter effects which are the dominant time-varying WFE, and 2) it mitigates the effects of misalignment and slow drifts. Moreover, it simplifies image reconstruction through mode-based imaging into a deconvolution problem. In the following subsections, we apply this technique to our $\beta$ CMi data.

\subsection{Response map construction from observations}\label{ssec:responsemap} 

The folllowing steps were used to construct PL response maps $M(\lambda,i)$ from the raw FIRST-PL and PSF camera data.

\begin{enumerate}
    \item Timestamp matching. First we matched timestamps between the two cameras. Since FIRST-PL camera ran slower than the PSF camera, we used FIRST-PL camera as a reference. We summed the dark-subtracted PSF camera frames to match the timestamps of the FIRST-PL camera, applying fractional multipliers when a frame fell between the two frames of the FIRST-PL camera.
    \item PSF peak determination. The peak location of the PSF in each synchronized frame was then determined by fitting a paraboloid function within a small window around the brightest pixel. This provided a sub-pixel-precision measurement of the PSF location.
    \item PL frame binning. Next, the raw PL frames were binned into a $15\times15$ grid spanning 48.6\,mas field of view (FoV), based on these peak locations (Fig. \ref{fig:method}b). We used 16.2 mas/pixel for the plate scale of the PSF camera. 
    This grid resolution and FoV were chosen as an empirical trade-off between maintaining sufficient angular resolution (lost with a coarser grid) and ensuring a high enough signal-to-noise ratio in each bin. In general, the optimal grid choice depends on the WFE conditions.
    \item Spectral extraction. We extracted spectra from each bin, resulting in $15\times15$ resolution response map for each port at every wavelength channel: $M(x,y;\lambda,i)$. See \S\ref{sec:a:reduction} for details on spectral extraction steps for FIRST-PL.
    \item Normalization. Finally, we normalized the maps such that their total sum across $15^2$ pixels equals unity, for each port and each wavelength channel; $\sum_{x,y} M(x,y;\lambda,i)=1$. This step decouples variation of spectral shapes from the variation of response maps across wavelengths.
\end{enumerate}

The resulting response maps of ports 1, 2, and 38 are shown in Fig. \ref{fig:method}d. The maps for different ports exhibit distinct morphologies, as each port responds differently to the wavefront (different linear combinations of the fiber spatial modes). See \S\ref{sec:a:fisher} for discussion on response maps in the Fisher information framework.

\subsection{Modeling reference response maps from the continuum}

To measure the reference response maps ($M_0$), a point-source calibrator may be observed separately. However, in practice, this is not trivial because from lab to on-sky, and from target to target, repeatable alignment within sub-mas precision is challenging, and $M_0$ can change based on upstream system alignment and AO-residual WFE properties (\S\ref{ssec:a:considerations}); we leave developing techniques using calibrator observations for future work. For $\beta$ CMi, we can adopt a spectral-differential approach, as it is nearly point-source-like in the visible continuum ($M_{\rm measured} \approx M_0$), dominated by stellar flux, allowing $M_0$ for the H$\alpha$ range ($|v|<500$~km s$^{-1}$) to be interpolated using adjacent wavelengths. The typical scale of lantern chromaticity is much broader than the width of the H$\alpha$ emission line, making such approach valid \citep{kim24_sa, kim24_3PL}. 

We model each port's response map at a given wavelength with a smooth, 9-th order bivariate polynomial in $x$ and $y$. We then fit the wavelength dependence of each spatial-polynomial coefficient using one-dimensional, 9-th order polynomial in $\lambda$. The example models along the wavelength axis with fixed ($x$, $y$) are shown as red lines in Fig. \ref{fig:method}d, and those in spatial axis with fixed $\lambda$ are displayed in Fig. \ref{fig:method}e. 

\begin{figure}[hbt!]
\centering
\includegraphics[width=1\columnwidth]{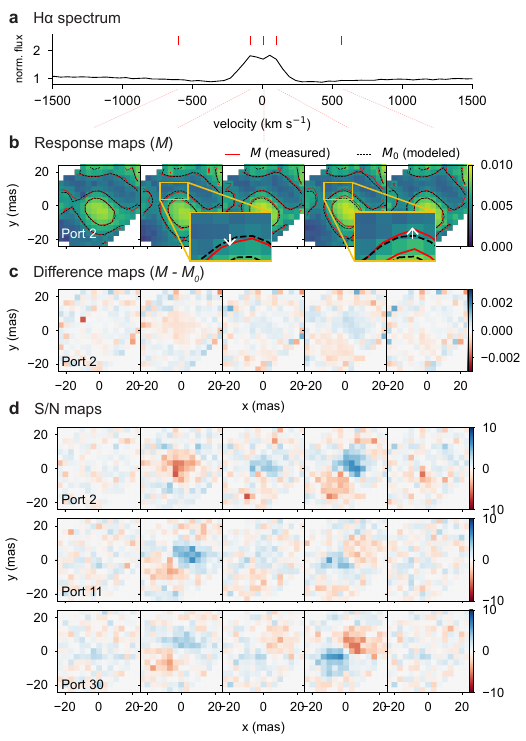}
\caption{
{Spectral-differential spatial signal detection.} 
{\bf a.} H$\alpha$ spectrum of $\beta$ CMi observed on 2025 February 11, summed over all the 38 traces. 
{\bf b.} Response maps (i.e., Fig. \ref{fig:method}e) of selected velocity channels, of one of the ports. Red solid contours and black dashed contours denote the same contour levels for the measured response maps and the modeled reference response maps. Small shifts are observed between the two contours as marked with white arrows in the zoom-in images, indicating photocenter shifts. 
{\bf c.} Response maps subtracted by the model reference response maps (difference maps). Deviations from zero indicate that the astronomical scene departs from an unresolved central point source. The positive and negative structures indicate the shift.
{\bf d.} Signal-to-noise (S/N) maps of three ports, which correspond to the difference maps divided by the error maps. The spatial signals are clearly seen in the H$\alpha$ wavelength range. 
}\label{fig:SNmap}
\end{figure}

\section{Results}\label{ssec:results} 

\begin{figure*}[hbt!]
    \centering
    \includegraphics[width=1.0\textwidth]{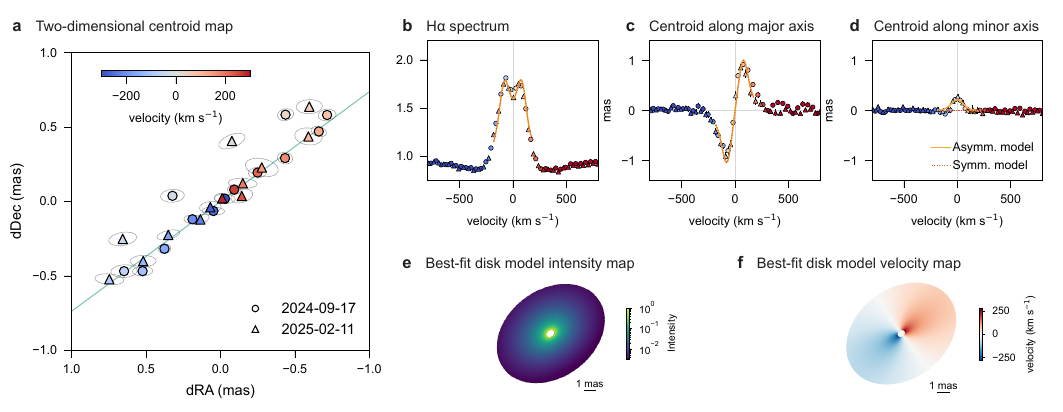}
    \caption{
    {Photocenter analysis of the H$\alpha$ decretion disk around $\beta$ CMi.} 
    {\bf a.} Two-dimensional photocenter positions for each spectral channel. Colors indicate the Doppler velocities, and gray ellipses show the 1\,$\sigma$ uncertainties. The green line denotes the derived disk position angle, 126$^{\circ}$.
    {\bf b.} H$\alpha$ spectrum of $\beta$ CMi (sum of the all 38 ports), showing double-peaked line profile. The solid and dotted lines indicate the best-fit Keplerian velocity disk models with asymmetric and symmetric power-law intensity distributions. 
    {\bf c, d.} Photocenter positions for each spectral channel, projected along the disk's major and minor axes, respectively. Photocenter measurements from two separate observations (circular and triangular markers) show strong agreement, demonstrating the repeatability of the measurement. The photocenter shift along the major axis reveals disk rotation consistent with a Keplerian velocity field. Interestingly, at low velocities, a shift along the minor axis is observed, indicating near- and far-side asymmetry of the disk.
    {\bf e.} Best-fit intensity distribution of the asymmetric Keplerian model.
    {\bf f.} Best-fit velocity distribution of the asymmetric Keplerian model.
    }
    \label{fig:SA}
\end{figure*}

We detected spatial signals at H$\alpha$ in all the 38 outputs as noticeable deviations of $M$ from modeled $M_0$. Fig. \ref{fig:SNmap} shows an example. In panel b, the measured maps ($M$, red solid contours) are offset from the reference maps ($M_0$, black dashed contours) in the second and fourth wavelength channels that correspond to the two H$\alpha$ peaks, with opposite directions. The same offset pattern was observed in response maps of other ports as well (see animated Fig. \ref{fig:video} in \S\ref{sec:a:video}). This indicates that the photocenters at those wavelengths are shifted relative to the continuum level in opposite directions. 

To reconstruct the image in each wavelength channel, we used all the 38 response maps and solved the deconvolution problem (Fig. \ref{fig:method}f). The following subsections describe image reconstruction results; detailed description of the steps are presented in \S\ref{sec:a:imgrecon_details}.

\subsection{Spectroastrometric analysis}\label{ssec:SA}

We began with the simplest image model to describe measured intensity responses: a point-source shifted by ($x$, $y$) mas. Fig. \ref{fig:SA}a presents the best-fit photocenter positions for each wavelength (Doppler velocity) channel. Two features are evident. First, the photocenters of blue- and red-shifted emission align linearly along the derived disk major axis (indicated by the green line), consistent with Keplerian rotation in agreement with previous studies \citep{wheelwright12, kraus12} (panel c). This is the major signal, identified in Fig. \ref{fig:SNmap}. Second, a small photocenter shift (0.27\,mas) is observed perpendicular to this axis (panel d). Along this axis, one side is closer to the observer than the other, given the inclination angle of the disk ($\theta_{\rm incl}=43^{\circ}$\citep{klement15}). Therefore the photocenter shift indicates there is apparent asymmetry between near and far sides. This is particularly interesting because simple optically thin axisymmetric disk models (displayed as dotted lines) do not predict such a shift. Previous Very Large Telescope Interferometer (VLTI) long-baseline spectro-interferometric observations of the Br$\gamma$ line yielded only a marginal detection of photocenter shift along the minor axis \citep{kraus12}. Adding an azimuthal modulation parameter $m_\phi$ to the Keplerian disk model (\S\ref{ssec:a:diskmodel}) yields a better fit, shown as solid lines (see Table \ref{tab:disk} for the derived disk parameters). 

We conjecture that the observed near-far side asymmetry arises from an opacity effect. The inner midplane region of the disk is at least marginally optically thick to H$\alpha$ photons, which obscures the photons emitted from the material located behind the midplane from the observer's point of view. The material located between the midplane and the observer for the near and far sides have different radial distances to the central star. Therefore, depending on the disk's density and thermal structure (which is predicted to be complex and have nonlinear dependence on stellar and disk parameters \citep{carciofi06_HDUST1, suffak23_HDUST3}), one side may appear brighter than the other. Such an effect has been observed in protoplanetary disks with optically thick molecular emission lines at radio frequencies \citep{rosenfeld13}, but this has never been observed for inner regions of hot Be star disks, with the exception of a marginal detection reported by \cite{kraus12}. High-precision spectroastrometry enabled by PLs can probe inner disk opacity effects and place constraints on their density and thermal structures.

Next, we attempted to measure spectroastrometric full-width at half-maximum (FWHM) signals \citep{porter04, baines06, wheelwright10, mendigutia18}: the size of the light-emitting region in each wavelength channel by extending our point source model to a circular 2D Gaussian model, allowing the model to fit the broadening of the response maps. Fig. \ref{fig:imgrecon}a displays the estimated Gaussian FWHM of each velocity channel. We note a hint of enhanced FWHM at H$\alpha$ from both observations, but the spurious signals in the continuum lowers its significance. Although the extended emission signal is marginal in this study, the results underscore the potential of PLs for size estimation at scales $<\lambda/2D$.

\subsection{Nonparametric image reconstruction}\label{ssec:imgrecon}

\begin{figure}[hbt!]
    \centering    \includegraphics[width=1\columnwidth]{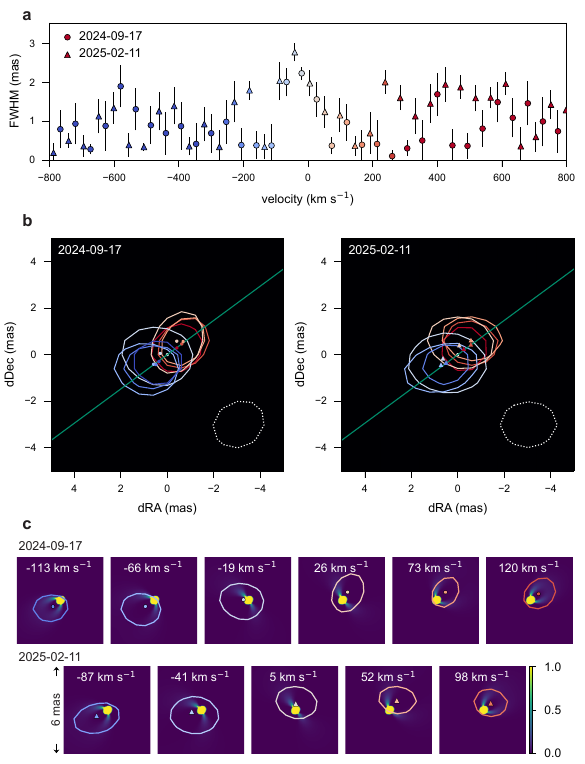}
    \caption{
    {Image reconstruction of the H$\alpha$ decretion disk around $\beta$ CMi.}
    {\bf a.} Measured circular 2D Gaussian FWHM, representing the spatial extent of the intensity distribution in each wavelength channel. 
    While there is an apparent FWHM enhancement around the H$\alpha$ center, it is of marginal statistical significance compared to the 1-2\,mas level of systematic fluctuations across the continuum.
    {\bf b.} The half-maximum level of reconstructed images as a function of wavelength ($|v|<180$km s$^{-1}$), for the two epochs. The colors indicate the velocity and has the same color scale as in panel a. The photocenter of each image is marked as circular and triangular symbols. The average images of the continuum level are shown in the lower right, analogous to the ``beam size", the instrumental width.
    {\bf c.} The half-maximum level of reconstructed images overlaid on best-fit asymmetric Keplerian model images for each wavelength channel from Fig. \ref{fig:SA}e, f. The reconstructed images show consistency with the model, tracing the shift in emission across wavelength --- both the rotation seen in Fig. \ref{fig:SA}f and the brightness asymmetry in Fig. \ref{fig:SA}e.
    }
    \label{fig:imgrecon}
\end{figure}

Extending further, we attempted non-parametric image reconstruction --- solving the deconvolution problem without any simplifying models. We used Markov chain Monte Carlo (MCMC) image reconstruction algorithm that originates from the interferometric image reconstruction problem \citep{macim, squeeze} to find the image that maximizes the likelihood given the response maps (see \S\ref{ssec:a:imgrecon} for details). Fig. \ref{fig:imgrecon}b displays the half-maximum contour of the reconstructed images, which show the photocenter shift and slight broadening around zero velocity as expected. Resolving finer details beyond centroids is challenging for this object, whose flux mostly comes from the central star and the disk emission is tightly confined around it (Fig. \ref{fig:imgrecon}c). Nevertheless, this shows the possibility of reconstructing the image using measured PL response maps.

\section{Discussion and outlook}\label{sec:discussion} 

This study presents the first on-sky science measurement made with a PL, achieving extreme spectral-differential photocenter measurement precision ($\sim$50\,$\mu$as) in the visible. We observed the late-type classical Be star $\beta$ CMi on two nights and, for the first time, resolved an H$\alpha$ brightness asymmetry between the near and far sides of its disk. We also measured its Keplerian rotation signature, which matches previous measurements. 

Moreover, this study introduces a practical mode-based imaging technique for astronomical observations. Mode-sorting imaging schemes have been extensively studied in theory (e.g., \cite{tsang16, rehacek17a, tsang18, tsang19, gessner20}) and demonstrated in stable laboratory conditions \citep{tham17, boucher20, rouviere24, santamaria25, deshler25}, but this is the first on-sky demonstration. Although robust measurement of higher-order image information than centroids was difficult for $\beta$ CMi in this study, greater performance could be achieved by alternate PL design. We used a standard 19-port photonic lantern that coherently mixes decomposed modes' complex amplitudes. The mixing introduces asymmetry of the response maps, lifting the sign degeneracy for photocenter displacements but compromises the sensitivity to higher-order image moments \citep{gessner20}. A mode-selective PL \citep{leon-saval14, xin24}, combined with simultaneous focal-plane images for response map reconstruction (i.e., capturing diverse responses), could potentially enable robust estimations of both the centroid and higher-order moments (see detailed discussion in \S\ref{ssec:a:FI}). 

PLs also offer practical benefits beyond the performance demonstrated here. As photonic devices, they are compact and cost effective compared to bulk optics, and can enable diverse instrument designs leveraging photonic technologies \citep{roadmap23}. Their single-mode outputs can feed existing diffraction-limited spectrometers \citep{betters14, lin21}, or even an on-chip spectrometer \citep{gatkine19}. They can feed a photonic circuit for further processing (such as interference of a pair of outputs to alter the transfer matrix or access mutual coherences) before feeding a spectrometer to optimize their performance \citep{lin23, kim24_imaging}. Their compactness can be suitable for space telescopes, where the absence of atmospheric turbulence could enable even greater super-resolution, especially on fainter targets that require longer integration times.

There are many more directions for improvement and expansion. We used spectral-differential techniques in this study, modeling the reference response maps from the target itself, but other calibration strategies could be developed independent of spectral-differential techniques to apply subdiffraction-limited measurement to more general objects. The knowledge of lantern transfer matrix as a function of wavelength, if well-characterized \citep{xin24, batarseh25}, could provide another route to process data. The design of the PL, such as the number of the modes and the properties of the transfer matrix, could be optimized for specific purposes \citep{lin23}. Lastly, as recently demonstrated on-sky, the PL output spectrum itself can be used for focal-plane wavefront sensing and control, which can correct for errors like the low-wind effect and enhance sensitivity by delivering a more stable wavefront correction during science observations \citep{lin25}.


\begin{acknowledgments}
Y.J.K thanks B. Hansen for helpful discussions. The authors would like to thank J. Laverty and S. Cunningham, the Subaru Telescope operators during the observations. This work is supported by the National Science Foundation under Grant Nos. 2109231, 2109232, 2308360, and 2308361. 
The development of SCExAO is supported by the Japan Society for the Promotion of Science (Grant-in-Aid for Research  No. 23340051, 26220704, 23103002, 19H00703, 19H00695 and 21H04998), the Subaru Telescope, the National Astronomical Observatory of Japan, the Astrobiology Center of the National Institutes of Natural Sciences, Japan, the Mt Cuba Foundation and the Heising-Simons Foundation. E.H. acknowledges funding support for the FIRST project by the French National Research Agency (ANR-21-CE31-0005), E.H. and S.L. acknowledge funding from the project ``Photonics" financed by the ANR program PEPR Origins (ANR-22-EXOR-0005). The authors wish to recognize and acknowledge the very significant cultural role and reverence that the summit of Maunakea has always had within the indigenous Hawaiian community, and are most fortunate to have the opportunity to conduct observations from this mountain.
\end{acknowledgments}

\begin{contribution}
M.P.F., N.J., S.L.-S., S.S., B.N., and O.G. developed the vision for the use of photonic lanterns for high angular resolution astronomy with SCExAO at the Subaru Telescope; 
Y.J.K. developed the concept and conceived the observing and image reconstruction strategy under the guidance of M.P.F. and in discussion with O.G., J. Lin, and Y.X.;
S. L.-S., S.Y., R.A.-C., B.N., and A.T. designed and fabricated the 19-port photonic lantern; 
M. Lallement, S.V.,  E.H., and S.L. designed and installed the FIRST-PL spectrometer at SCExAO;
S.V., M. Lallement, E.H., S.L., M.N. and J.S. designed and integrated the photonic lantern injection setup at SCExAO, with technical guidance of O.G., N.J., B.N., and M.P.F.;
O.G., V.D., and M. Lucas designed and implemented the SCExAO software for data acquisition and logging;
O.G., S.V., E.H., and S.L. selected the target for the observations;
S.V. and Y.J.K. acquired data with the FIRST-PL and the internal PSF camera during the observations;
O.G., J. Lozi, V.D., S.V., and M. Lucas operated the SCExAO and adaptive optics system during the observations.
M. Lucas acquired calibration data for the internal PSF camera;
Y.J.K. and M.P.F. developed strategies for the FIRST-PL detector calibration and spectral extraction and implemented the software; 
J. Lin, Y.J.K., Y.X., J. Lozi, and S.V. developed calibration and control software for experiments with photonic lanterns at SCExAO;
Y.J.K. performed daytime experiments with the FIRST-PL and the internal camera with assistance from J. Lozi, S.V., V.D., J. Lin, Y.X., and M. Lucas, created the response map modeling and image reconstruction software, led the data analysis and interpretation with guidance from M.P.F., and wrote the manuscript with input from all co-authors.



\end{contribution}

%
\facilities{Subaru Telescope (SCExAO)}

\software{
SciPy \citep{scipy},
NumPy \citep{numpy},
Matplotlib \citep{matplotlib}
          }


\appendix

\section{Observation details}\label{sec:a:obslog}

\begin{figure}[hbt!]
    \centering
    \includegraphics[width=1\columnwidth]{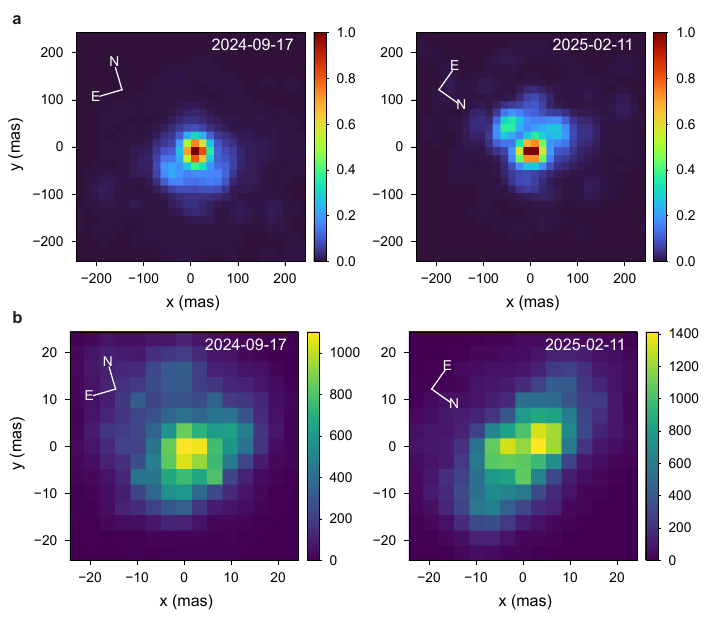}
    \caption{
    {Average PSF and PSF peak location histogram.}
    {\bf a.} Average $H$-band PSFs of the two observations. For both observations, the low-wind effect affected the PSF quality, splitting the core.
    {\bf b.} Distribution of PSF peak positions. The tip-tilt jitter in September 2024 is nearly circular, whereas in February 2025 it shows an elongated distribution.
    }
    \label{fig:psf_extended}
\end{figure}

The observation log for both nights is presented in Table \ref{tab:obslog}. Fig. \ref{fig:psf_extended} summarizes the PSF quality from our observations; the top panels display the time-averaged PSF, while the lower panels show a histogram of the instantaneous PSF peak locations.

The observations were conducted in pupil-tracking mode, in which an image rotator tracks the pupil while the sky field rotates in the PL injection plane (as well as the PSF camera). Given the small change in parallactic angle during the observation for both dates (Table \ref{tab:obslog}), we did not apply field-rotation correction to the data. In this mode, the image rotator angle can be used to determine the sky orientation, with an offset amount to be calibrated. We adopted a 2.7$^\circ$ position angle calibration offset for the PSF camera from earlier measurements; because this value has not been revised in recent years, it may introduce a systematic uncertainty of a few degrees into our position angle measurement.

\begin{deluxetable}{lcc}[hbt!]
\tabletypesize{\scriptsize}
\tablecaption{$\beta$ CMi Observation log\label{tab:obslog}}%
\tablehead{
\colhead{Observation date} & \colhead{2024-09-17} & \colhead{2025-02-11}
}
\startdata
Observation time (UT) & 14:55:17 -- 15:09:17 
 & 12:11:08 -- 12:19:18 \\
Parallactic angle (deg) & -70.96 -- -70.33 & 72.03 -- 72.11 \\
FIRST-PL camera DIT (ms) & 20 & 10 \\
PSF camera DIT (ms) & 5.3 & 2.7 \\
$H$-band avg. Strehl & 0.55 & 0.51 \\
RMS tip-tilt jitter (mas) & 16 & 13\\
\enddata
\end{deluxetable}

\section{Detector nonlinearity correction and spectral extraction}\label{sec:a:reduction}

After step 3 in \S\ref{ssec:responsemap}, we have $15\times15$ number of averaged PL frames, sorted by PSF spatial coordinates. Within each bin, we subtracted a dark frame from the raw frames, averaged them, and applied a detector nonlinearity correction to account for the qCMOS' low-count nonlinear response \citep{strakhov23}. To model the nonlinear behavior in each pixel, we performed a series of flat-field exposures using a halogen lamp with various integration times. For each pixel, we fit a simple linear model to its response at the highest count levels, then constructed an empirical correction curve that maps the raw counts back to the expected linear counts at all exposure levels. For the February 2025 dataset, we applied the full per-pixel nonlinearity correction. In contrast, no flat-field sequence existed for September 2024, and its spectral traces did not align with the flat spectrum we acquired later; therefore, we used the average nonlinearity curve to correct the global nonlinearity response of that dataset. Thus, we were not able to correct for the pixel-to-pixel variation of nonlinear response for the September 2024 data, but this was relatively less significant compared to the February 2025 data because the count levels were generally above the low-count regime where detector nonlinearity is severe.

From the nonlinearity-corrected averaged frame within each bin, the spectra were extracted by forward-modeling spectral traces on the detector. The models were constructed using empirical line-spread functions (LSFs) recorded with a Neon lamp. We selected 12 high signal-to-noise LSFs and interpolated between them to construct model LSFs at any wavelength. The 38 spectra were then extracted by performing a least-squares fit of nonlinearity-corrected average frames to these models. The wavelength solution for each port was constructed using the same Neon lamp data. The extracted spectra were interpolated in wavelength space to align with the wavelength grid of the first trace. For H$\alpha$ spectrum, we applied barycentric velocity correction and subtracted $\beta$ CMi's radial velocity of 22~km s$^{-1}$.

The variances of the response maps were estimated by bootstrap resampling. We resampled the frames within each bin with replacement, averaged them, applied nonlinearity correction, and applied steps 4-5 in \S\ref{ssec:responsemap} to generate a bootstrap sample of response maps. We repeated this process 50 times and calculated the variance map from the variance of the samples.

\section{Spectral-differential spatial signal detection on principal components of normalized spectrum}\label{sec:a:PCA}

\begin{figure*}[hbt!]
    \centering
    \includegraphics[width=1\textwidth]{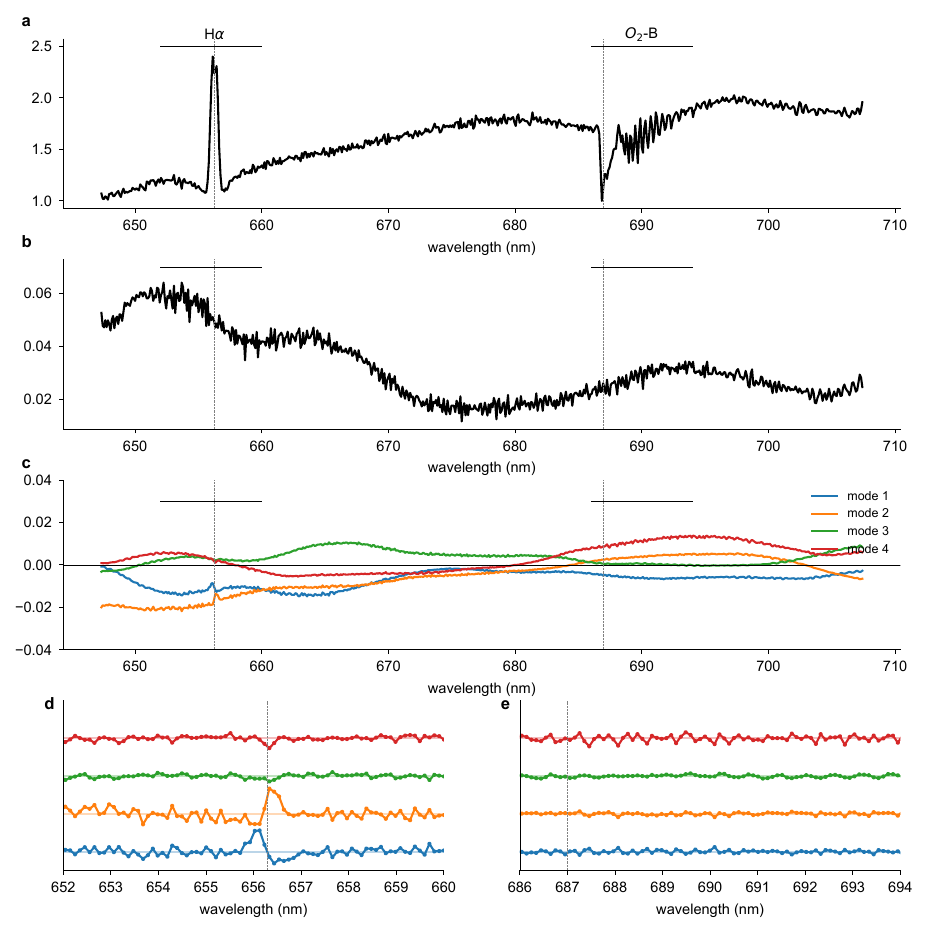}
    \caption{
    {Principal component analysis of the normalized spectrum.}
    {\bf a.} Spectrum of $\beta$ CMi observed on September 2024, showing double-peaked H$\alpha$ line from the decretion disk and oxygen B-band absorption due to the Earth's atmosphere.
    {\bf b.} Average normalized spectrum of one of the output ports. Spectral-differential spatial signals are not detected in the average spectrum, due to signal dilution by time-varying WFEs and fixed-pattern noise by detector pixel-to-pixel variation. 
    {\bf c.} The first four principal components of the time-series normalized spectrum. 
    {\bf d.} Zoomed-in cutouts of panel c around H$\alpha$. The low-frequency component was filtered and the four components were shifted vertically for visualization. Emission line-scale features are detected in the principal components, implying the presence of spectral-differential spatial signals. 
    {\bf e.} Same as panel d, but for the oxygen B-band. No line-scale features are detected.}
    \label{fig:pca}
\end{figure*}

PLs encode wavefront information in the relative intensities of their output ports \citep{norris20, lin22}. When the outputs are spectrally dispersed, the equivalent quantity is the normalized spectrum, defined as the spectrum of each port divided by the summed spectrum across all the ports (Fig. \ref{fig:pca}b). PLs' intrinsic chromaticity makes normalized spectra and their response to wavefront variation vary slowly with wavelength \citep{kim24_sa, kim24_3PL, lin25}. For a point-source unresolved at all wavelengths, the normalized spectrum depends only on wavefront variations and not on the source's intrinsic spectrum. If the source's morphology changes across a spectral line, the normalized spectrum and its response to wavefront variation should exhibit distinctive line-scale features. For instance, if one member of a binary has an emission line, the intensity distribution across the ports for one star and the other are different, and that behavior would cause different levels in normalized spectrum for the emission line and the continuum \citep{kim24_sa}. 

We first examined the normalized spectrum over a broad wavelength range. We extracted spectra from each dark-subtracted frame as in \S\ref{sec:a:reduction} (but without nonlinearity correction; nonlinearity correction on low S/N individual frames can introduce bias) and computed the normalized spectrum. The averaged normalized spectrum shows no clear feature in H$\alpha$ (Fig. \ref{fig:pca}b); instead, it is dominated by high-frequency features, which are mainly artifacts due to pixel-to-pixel variation of the detector nonlinearity. Tip-tilt jitter further dilutes any spatial signals in the average normalized spectrum \citep{kim24_sa}. Thus, we investigated frame-to-frame variance in the normalized spectrum, which is more robust to fixed pattern artifacts, and more immune to dilution by tip-tilt jitter. 

We flattened the normalized spectrum to an array with length of 38 $\times N_\lambda$ for each frame. Then we performed principal component analysis (PCA) on the time-series normalized spectrum array. The principal components should reflect the modes of variation in PLs' response to the time-varying wavefront. Fig. \ref{fig:pca}c displays the first four principal components, for one of the PL output ports. The low-order variation in the principal components correspond to the PLs' intrinsic chromaticity, which makes the response vary slowly across the wavelength \citep{kim24_sa, kim24_3PL}. In H$\alpha$, emission-line scale features are detected, dominantly in modes 1 and 2 (panel d). We have detected similar signals in most other ports as well. From this, we concluded that spatial signatures are present in H$\alpha$ but not in other wavelengths, and proceeded forward to reconstruct the intensity distribution using the response maps. 

\section{Response map animation}\label{sec:a:video}

\begin{figure*}[hbt!]
\begin{interactive}{animation}{efig3_anim.mp4}
    \centering
    \includegraphics[width=1\textwidth]{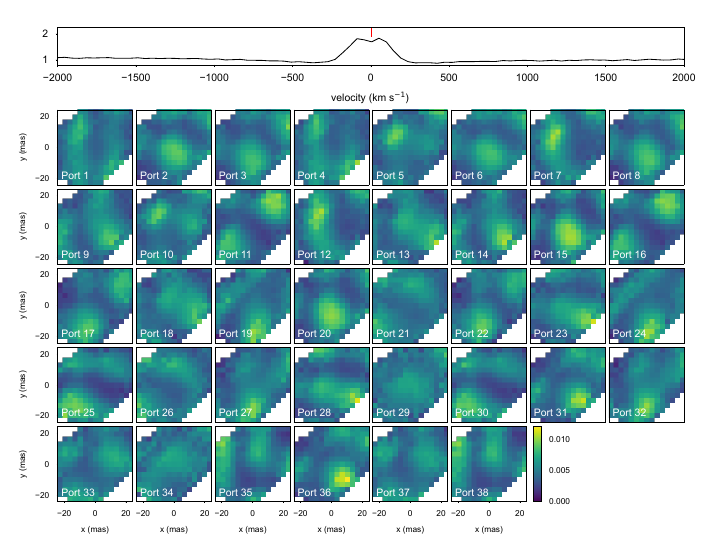}
\end{interactive}
    \caption{Same as Fig. \ref{fig:SNmap}a, b but for all the 38 ports at the H$\alpha$ center. 
    The full figure is available as an animation of 3\,s duration. In the animation, a red vertical marker, annoated on the top panel's spectrum, sweeps through velocity channels from -1000\,km s$^{-1}$ to 1000\,km s$^{-1}$. The bottom 38 panels display the response maps corresponding to the velocity channel indicated by the marker.
    The animation shows a coherent shift of all the response maps across the H$\alpha$ emission line, revealing the photocenter displacement signal. \label{fig:video}}
\end{figure*}

For a complete visualization of the multiple response maps as a function of wavelength, we present the animated maps in Fig. \ref{fig:video}.

\section{Details on image reconstruction method}\label{sec:a:imgrecon_details}

\subsection{Point source and Gaussian model fitting}\label{ssec:a:modelfitting}

Response maps of any intensity distribution can be calculated by convolving model intensity distribution with the model response maps for a central point source $M_0$ (Equation \ref{eq:model}). We used a point-source model (which is essentially translational shift) for the photocenter analysis and two-dimensional circular Gaussian model for the FWHM estimation in Section \ref{ssec:SA}. 

For each wavelength, we computed $\chi^2$ to quantify the goodness of fit between the model and the measured response maps. Specifically, we summed over all ports and spatial bins the squared residuals divided by the variance maps. The uncertainties in our parameter estimates were quantified via bootstrap resampling of the 38 port spectra: for each of 50 trials, we drew 38 spectral traces at random with replacement from the full port set, refit the model, and recorded the resulting parameters. The standard deviation of those 50 bootstrap-derived parameter sets was taken as the 1$\sigma$ error on each parameter. 

\subsection{Non-parametric image reconstruction}\label{ssec:a:imgrecon}

The representation of model intensity distribution has two parameters for the point source model and three parameters for the Gaussian model in the previous subsection. Generalizing this, we can think of an image model represented in pixel basis of regular grid. Then the image reconstruction problem becomes finding the most probable flux in each pixel, given measured response maps. 

We used a simulated annealing with the Metropolis sampler as in {\tt MACIM} \citep{macim} and {\tt SQUEEZE} \citep{squeeze}. In this framework, an image is represented by a collection of flux elements, existing on a regular grid of image pixels. This imposes positivity of the image. Initially, the flux elements are randomly spread over the image grid. Then each step, a tentative move of each flux element to a random adjacent pixel is proposed with an acceptance probability defined by the Metropolis formula modified by the annealing temperature defined as
\begin{equation}
    T_{j+1} = T_j + \frac{1}{\tau}\left(\chi^2_{r,j} - \gamma T_j\right)\left(1 - \frac{\chi^2_t}{\chi^2_{r,j}}\right)
\end{equation} 
where $T_j$ is the annealing temperature at $j$th step, $\gamma$ is the temperature decay coefficient, 
$\chi^2_{r,j}$ is the reduced $\chi^2$ ($\chi^2$ divided by the number of degrees of freedom) at $j$th step, $\chi^2_t$ is the target $\chi^2_r$ during convergence, and $\tau$ is the timescale of temperature changes \citep{macim, squeeze}. We find that using 50--100 flux elements on $33\times33$ image grid with a 20\,mas field of view achieves good convergence for our problem without requiring additional regularization or an informative prior. The reconstructed images are sensitive to the choice of $\gamma$: as $\gamma$ increases, the annealing temperature during convergence --- and hence the acceptance probability --- rises, yielding systematically broader posterior images. Experimenting with various values of $\gamma$ and with different set of data generated by bootstrap resampling (38 subset with replacement from 38 spectral traces), we found that low $\gamma$ leads to underfitting (loss of detail), while high $\gamma$ results in overfitting (spurious features). We chose $\gamma = 10^4$ for the images displayed in Fig. \ref{fig:imgrecon}, which overall resulted in slightly more extended images than the best-fit Gaussian models.

\section{Simple disk model fitting}\label{ssec:a:diskmodel}

Following \cite{kraus12}, we adopt a Keplerian disk model in which rotation velocity at radial distance $r$ is described as $v(r) = \sqrt{GM_*/r}$ and the surface density follows a power-law, $\Sigma(r) \propto r^{q}$, from the stellar equator out to a truncation radius $R_{\rm out}$. This represents a geometrically and optically thin disk, with $q=-2$ corresponding to the isothermal case. We fixed the stellar equatorial and polar radii to 0.34\,mas and 0.23\,mas, respectively \citep{klement15}. In addition, to model the asymmetry along the disk's minor axis, we add an azimuthal modulation parameter $0\leq m_\phi \leq 1$ describing asymmetric intensity distribution:
\begin{equation}
    \Sigma(r,\phi) \propto r^q\left(1 + \frac{m_\phi}{2}\cos{\phi}\right).
\end{equation}

We removed the central star's contribution in the observed quantities (H$\alpha$ spectrum and centroid shifts; Fig. \ref{fig:SA}b,c,d) before fitting them to the model, assuming that the star is unresolved. Although the star's rapid rotation can produce small photocenter shifts across the H$\alpha$ absorption line, these effects are negligible compared to the much stronger signals from the disk given the star's much smaller angular extent and weaker absorption depth compared to the emission line strength. We modeled the stellar flux contribution to the H$\alpha$ spectrum in each wavelength ($f_\lambda$) by fitting the absorption wings in the measured spectrum to the {\tt PHOENIX} stellar models ($T_{\rm eff} = 12,000$\,K, $\log{g} = 3.5$) \citep{phoenix}. The disk-only H$\alpha$ emission spectrum was estimated by multiplying $(1-f_\lambda)$ to the observed spectrum and the disk-only centroid shifts were derived by dividing the observed shifts by $(1-f_\lambda)$. Fitting was performed over velocity channels of $|v|< 200$\,km s$^{-1}$.

Table \ref{tab:disk} lists the retrieved parameters. The solid lines in upper panels of Fig. \ref{fig:SA} represent the best-fit disk model with the azimuthal modulation parameter, while the dotted lines correspond to the symmetric case, $m_\phi = 0$. The intensity and velocity distribution of the best-fit disk model are shown in lower panels of Fig. \ref{fig:SA}. The retrieved power-law index $q=-2.0\pm0.1$ agrees well with the isothermal density profile, and the outer disk radius $R_{\rm out} = 5.2\pm0.3$~mas is close to the value found by \cite{kraus12} in Br$\gamma$, $5.8\pm0.2$~mas. The disk rotation velocity $V_{\rm rot} = 430\pm30$\,km s$^{-1}$ and the inclination angle $\theta_{\rm incl}=42\pm3^\circ$ also match values reported in previous work (e.g., \cite{klement15}). We find that the azimuthal modulation parameter $m_\phi = 0.50\pm0.05$ best describes the photocenter shift along the minor axis with our simple model, but reproducing the detailed profile requires higher-order terms, since the observed shift is narrower than predicted by a pure first-order modulation.

\begin{deluxetable*}{lccc}[bt!]
\centering
\tabletypesize{\scriptsize}
\tablecaption{Retrieved Keplerian disk model parameters\label{tab:disk}}
\tablehead{
\colhead{Parameter} & & \colhead{Value} &
}
\startdata
Disk rotation velocity at stellar equatorial radius &$V_{\rm rot}$  & $430\pm30$ & (km s$^{-1}$)\\
Disk inclination angle & $\theta_{\rm incl}$ &  $42\pm3$  & (deg)\\
Disk position angle  & PA & $126$\tablenotemark{a}& (deg)\\ Disk outer radius & $R_{\rm out}$  & $5.2\pm0.3$& (mas) \\
Radial power-law index & $q$ &  $-2.0\pm0.1$ & \\
Azimuthal modulation parameter & $m_\phi$ &   $0.50\pm0.05$ &\\
\enddata
\centering
\tablenotetext{a}{Subject to systematic uncertainty of the image rotator position angle offset.}
\end{deluxetable*}

\section{Classical Fisher information analysis of photonic lantern observables}\label{sec:a:fisher}

Imaging via a PL is closely related to the spatial mode demultiplexing (SPADE) technique \citep{tsang16}, which decomposes focal-plane field into desired mode bases for quantum-optimal parameter estimation in the subdiffraction regime. 
Fisher information is typically used as a metric, which sets general lower bounds on the error of parameter estimation. 
Considering the problem of estimating the separation between two equal point sources, for any real symmetric PSF, modal decomposition on definite even/odd modes achieves constant Fisher information even when the separation approaches zero \citep{tsang16, rehacek17a}. This can be generalized to higher-order image moment estimation, achieving enhancement in Fisher information over focal-plane intensity imaging \citep{tsang18, tsang19}.
Due to symmetry of the modes, intensity measurements in each mode does not provide information on odd image moments; odd moments can be assessed by interference of a pair of modes \citep{yang16,tsang18}. Although extensively studied in theory, experimental implementations and demonstrations of SPADE remain mostly limited to simple two point source separation estimation \citep{tham17, boucher20, rouviere24, santamaria25}.

A mode-selective PL could be a direct implementation of SPADE, as it maps each LP mode amplitude to each PL output. A standard PL, on the other hand, does not have symmetry required for SPADE measurements; rather, it can be viewed as SPADE with significant crosstalk \citep{gessner20}, determined by the transfer matrix.

In \S\ref{ssec:a:observables} we provide a detailed mathematical framework of PL observables; in \S\ref{ssec:a:FI} we compute Fisher information for representative parameter estimation problems; in \S\ref{ssec:a:considerations} we provide notes on practical measurement requirements under on-sky conditions.

\subsection{Photonic lantern observables}\label{ssec:a:observables}

Let us consider an $N$-port photonic lantern (PL) that supports $N$ number of modes of propagation. When its multimode entrance is placed at the focal plane, the focal-plane field up to $N$ fiber modes can couple into the PL. Assuming a circular symmetric step-index fiber, these modes correspond to LP modes. Writing $a_j$ as the overlap between the focal-plane field $\psi$ and $j$-th fiber mode $\phi_j$:
\begin{equation}\label{eq:overlap}
    a_j \equiv \langle \psi\mid \phi_j\rangle,
\end{equation}
this corresponds to the $j$-th mode (complex-valued) amplitude of the input field. The single-mode outputs are linear combinations of these mode amplitudes, 
\begin{equation}\label{eq:transfer_matrix}
    {E}_i = \sum_{j=1}^N U_{ij}a_j
\end{equation}
where $E_i$ is the complex amplitude of $i$-th port and $U$ is the complex-valued transfer matrix \citep{lin22, kim24_3PL}. The transfer matrix is determined by the design and construction of the PL \citep{lin23}. For instance, the transfer matrix of an ideal mode-selective PL \citep{leon-saval14} is identity, mapping each LP mode amplitude to each output. A standard lantern, a kind used in this study, generally has significant off-diagonal components, mixing all the mode amplitudes.

By imaging PL outputs on a detector, one measures intensities in each port,
\begin{equation}
    I_i = |E_i|^2.
\end{equation}
The PL outputs may feed a single-mode fiber-fed spectrometer to measure intensities spectrally, which is done in this study. While directly measuring each output's intensity is the simplest observable, one can also interfere pairs of outputs to access their mutual coherence or alter the effective transfer matrix \citep{lin23, kim24_imaging}. This can be realized by feeding PL outputs to a beam combiner chip for example.

The intensity measurement can be made as scanning the focal-plane field in $x$ and $y$ across the PL's multimode input tip, which we term in the paper the response map: $M(x,y;i,\lambda)$. In practical observing conditions, the response maps can help calibrate misalignment and diagnose WFE effects \citep{kim24_3PL}, as will be discussed in \S\ref{ssec:a:considerations}.

\subsection{Classical Fisher information of PL observables for subdiffraction imaging}\label{ssec:a:FI}



We compare the mode-selective PL and standard PL for two cases: point-source localization and separation measurement of two equal sources.

\begin{figure*}[hbt!]
    \centering
    \includegraphics[width=1\linewidth]{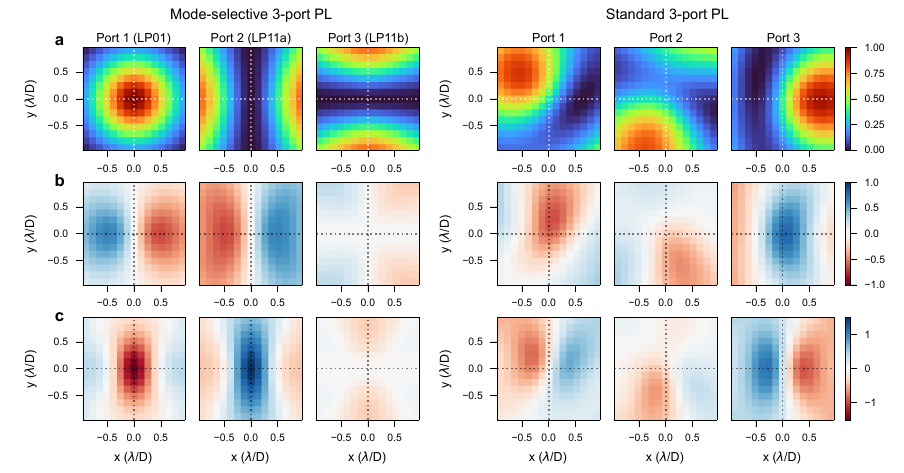}
    \caption{
    {Simulated response maps and their derivatives for standard and mode-selective 3-port PLs}
    {\bf a.} Example response maps ($M'_{0i}$) of mode-selective (left) and standard (right) 3-port PLs. A random unitary matrix is used to simulate standard PL responses. The maps are normalized such that $\sum_{i=1}^{3} M'_{0i}=1$.
    {\bf b.} Gradients of $M'_{0i}$ along $x$. While the mode-selective PL exhibits zeros at the center, the standard PL has substantial gradients.
    {\bf c.} Second-order derivatives of $M'_{0i}$ along $x$. Large second-order derivative and zero intensity at (0, 0) for the LP$_{\rm 11a}$ port of the mode-selective PL suggests nonzero Fisher information at the small separation limit (Eq. \ref{eq:FI_separation}).
    }
    \label{fig:simulated_couplingmaps}
\end{figure*}

Let us assume an ideal PL with 100\% throughput aligned at $\mathbf{X}= (0,0)$ and a perfect PSF matching the fundamental mode of the PL. Conditioned on a photon detection event through the lantern, the probability of detecting the photon in $i$-th port can be written as $P_i$ where $\sum_{i=1}^N P_i = 1$. The Fisher information matrix for estimation of parameters $\mathbf{\theta}$ from measurement of $P_i$ given $N_{\rm phot}$ number of detected photons is then
\begin{equation}
    [\mathcal{I}(\mathbf{\theta})]_{j,k} = N_{\rm phot}\sum_{i=1}^N \frac{1}{P_i} \frac{\partial P_i}{\partial \theta_j} \frac{\partial P_i}{\partial \theta_k}.
\end{equation}

If the input scene is a single point source located at $\mathbf{X} = (x,y)$, $P_i(\mathbf{X})$ corresponds to the reference response map defined in this work, but normalized over the ports instead of the image plane. We write the reference response maps of $i$-th port with this different normalization as $M'_{0i}(\mathbf{X})$ where $\sum_{i=1}^{N}M'_{0i}(\mathbf{X}) = 1$. Or, written in terms of $I_{0i}(\mathbf{X})$, which refers to intensities in each port for a point source located at $\mathbf{X}$,
\begin{equation}
    M'_{0i}(\mathbf{X}) = \frac{I_{0i}(\mathbf{X})}{\sum_{i=1}^N I_{0i}(\mathbf{X})}.
\end{equation}
This allows us to relate the PL response maps to Fisher information for parameter estimation. 
Fig. \ref{fig:simulated_couplingmaps}a shows simulated response maps ($M'_{0i}$) of a mode-selective 3-port PL and an example standard 3-port PL for illustration. 

\subsubsection{Estimation of single point source position}

If the input scene consists of a single point source at an unknown position $\mathbf{W}$, the probability of detecting the photon in $i$-th port is then
\begin{equation}
    P_i = M'_{0i}(\mathbf{W})
\end{equation}
and the Fisher information matrix for estimation of $\mathbf{W}$ becomes
\begin{equation}
    [\mathcal{I}(\mathbf{W})]_{j,k} = N_{\rm phot} \sum_{i=1}^N \frac{1}{M'_{0i}(\mathbf{W})} \frac{\partial M'_{0i}}{\partial X_j}\Big\vert_{\mathbf{W}}\frac{\partial M'_{0i}}{\partial X_k}\Big\vert_{\mathbf{W}}.
\end{equation}
This implies that the gradient of the response maps (Fig. \ref{fig:simulated_couplingmaps}b) at the position of the object is important for localizing the object. In the special case of $M'_{0i}(\mathbf{W}) \rightarrow 0$, expansion up to second-order around $\mathbf{W}$ ensures convergence to nonzero $\mathcal{I(\mathbf{W})}$. An ideal mode-selective PL, however, has sign ambiguity in source localization due to symmetry: $M'_{0i}(\mathbf{W}) = M'_{0i}(-\mathbf{W})$ \citep{xin22, xin24}. The ambiguity may be lifted by interference of modes \citep{tsang18} or taking measurements at multiple positions. In contrast, a standard PL generally has nonzero gradient terms and substantial asymmetry, crucial for positional measurement \citep{kim24_3PL}.

\subsubsection{Estimation of separation between two equally bright point sources} 

Now we consider the classic problem of estimating the separation between two equally bright point sources. Note that the separation measurement problem is closely related to the second image moment or object size estimation problem \citep{tsang18, dutton19}.
Let us consider a scene that consists of two incoherent point sources with equal brightness. Writing the separation vector as $\mathbf{S}$ and the centroid position as $\mathbf{W}$, $P_i$ is then
\begin{equation}
    P_i = \frac{1}{2}\left[M'_{0i}\left(\mathbf{W} + \frac{1}{2}\mathbf{S}\right) + M'_{0i}\left(\mathbf{W} - \frac{1}{2}\mathbf{S}\right)\right].
\end{equation}
At the small separation limit, Taylor expansion around $\mathbf{W}$ up to second order leads to
\begin{equation}
    P_i \approx M'_{0i}(\mathbf{W}) + \frac{1}{8}\mathbf{S}^{\top} \mathcal{H}_i(\mathbf{W})\,\mathbf{S}
\end{equation}
where $\mathcal{H}_i(\mathbf{W})$ is the Hessian matrix of $M'_{0i}$ evaluated at $\mathbf{W}$, $[\mathcal{H}_i(\mathbf{W})]_{j,k}=\frac{\partial^2 M'_{0i}}{\partial X_j \partial X_k}\big\vert_{\mathbf{W}}$.
Further assuming that the centroid position $\mathbf{W}$ is known but $\mathbf{S}$ is unknown, the Fisher information matrix for estimation of $\mathbf{S}$ is
\begin{equation}\label{eq:FI_separation}
    [\mathcal{I}(\mathbf{S})]_{j,k} \approx N_{\rm phot} \sum_{i=1}^N \frac{1}{16} \frac{\bigl(\mathcal{H}_i(\mathbf{W})\,\mathbf{S} \bigr)_j\bigl(\mathcal{H}_i(\mathbf{W})\,\mathbf{S} \bigr)_k}{M'_{0i} (\mathbf{W}) + \frac{1}{8}\mathbf{S}^{\top} \mathcal{H}_i(\mathbf{W})\,\mathbf{S}} .
\end{equation}
For the case of mode-selective PL, when perfectly aligned at the centroid ($\mathbf{W}=0$), constant Fisher information in the $\mathbf{S} \rightarrow 0$ limit is achieved in the LP$_{\rm 11}$ mode, as in \cite{tsang16}. However, unless any of the ports exhibit zero intensities at $\mathbf{W}$ ($M'_{0i}(\mathbf{W})=0$), due to misalignment or crosstalk, the Fisher information falls to zero quadratically as $\mathbf{S} \rightarrow 0$. The quadratic fall-off starts at shorter separations as $M'_{0i}(\mathbf{W})$ gets closer to zero. Thus, while a mode-selective PL can offer a clear advantage over direct imaging for second-moment estimation, a standard PL with substantial mode mixing may not have a comparable advantage. The overall level of the Fisher information is governed by the local second-order derivatives of the response maps (Fig. \ref{fig:simulated_couplingmaps}c), corresponding to the width of the point spread function (PSF) of direct imaging. These derivatives govern the sharpness of the clumpy features of the response maps. The sharpness of these features may be modified by adjusting the focal ratio, but at the expense of reduced coupling efficiency.

\subsection{On-sky experimental considerations}\label{ssec:a:considerations}

In stable laboratory conditions, it is sufficient to measure $P_i$ at the target and use pre-determined $M'_{0i}(\mathbf{X})$ taken at a calibration source to characterize the mode-sorting device (e.g., \cite{rouviere24}). A good number of samplings in $\mathbf{X}$ is required to measure derivatives of $M'_{0i}$ at the location of the target. However, in ground-based astronomical observation conditions, $M'_{0i}$ taken in laboratory cannot be used directly to calibrate measurements on-sky. Rapid fluctuations from atmospheric turbulence and telescope vibration as well as uncorrected static aberrations alter $M'_{0i}$ \citep{kim24_3PL}. $M'_{0i}$ can change even during the observation as conditions change. Thus, $M'_{0i}$ should be measured on-sky, with as similar wavefront conditions as possible as the target.

The technique we introduce in the paper constructs response maps on-sky, leveraging the natural jitter. Further using the fact that the target is point-source-like in the continuum and the PL transfer matrix varies slowly with wavelength \citep{kim24_sa}, $M'_{0i}$ in the H$\alpha$ wavelength range could be modeled. Thus, the reference and target response maps could be observed simultaneously. 

For targets without such spectral-differential features, a calibrator may be observed and used as $M'_{0i}$. Even if a calibrator is observed separately, it is likely that response maps are needed on the target ($M'_{i}$) as well, because misalignments must be robustly determined from the data. In practical on-sky conditions, it is generally difficult to align the target at the same position as the calibrator within milliarcsecond level precision. Especially for a mode-selective PL, which is optimal for second moment measurement but has degeneracy in positional measurement, would benefit from diverse measurements at different positions that $M'_i$ provides.

\bibliography{sample701}{}
\bibliographystyle{aasjournalv7}



\end{document}